\documentstyle[11pt,newpasp,twoside,epsf]{article}
\def\edcomment#1{\iffalse\marginpar{\raggedright\sl#1\/}\else\relax\fi}
\reversemarginpar

\begin{document}
\title{Rebrightening  in GRBs by Precessing  Jet}
\author{Daniele Fargion}
\affil{Physics Department and INFN, Rome University, "La
Sapienza",Roma,Italy}

\begin{abstract}
A  clear gamma polarization in the $\gamma$ signals from
GRB$021206$ probes  the presence of a very thin collimated jet
(opening angle $\Delta\theta \leq 0.6^o$;
$\frac{\Delta\Omega}{\Omega}\leq 2.5 \cdot 10^{-5} $) in Gamma Ray
Burst, GRBs. The  last and well proved GRB$030329$/SN$2003$dh
time and space coincidence confirms definitively  the earliest
GRB$980425$/SN$1998$bw connection among GRB and Supernova. The
apparent  extreme GRBs  luminosity is just the extreme beamed
blazing gamma Jet observed in axis during a Supernova birth. The
maximal isotropic SN power, $\dot{E}_{SN}$$ \simeq 10^{45}$ erg
$s^{-1}$, because of its probable energy equi-partition, should be
collimated even into an even thinner jet
$\frac{\Delta\Omega}{\Omega}\leq 10^{-8}$, in order to explain at
once the apparent observed maximal GRBs output,
$\dot{E}_{GRB}\simeq$ $\dot{E}_{SN}$$\frac{\Omega}{\Delta\Omega}$
$ \simeq 10^{53}$ erg $s^{-1}$. Consequently one-shoot thin Jet
GRBs needs many more $\dot{N}_{GRBs} \simeq$
$\frac{\Omega}{\Delta\Omega}\cdot \dot{N}_{Fireball}$ event rate
than any spread isotropic Fireballs. Such a rate exceed
 even the known Supernova one. To overcome the puzzle a persistent precessing
decaying  jet (life-time $\tau_{Jet}$ $ \geq 10^3$ $\tau_{GRB}$)
is compelling. The late relic GRBs  sources may be found in
compact SNRs core, as  NS or BH jets;  at much later epoch, their
lower power $\gamma$ jets may be within detectability only from
nearby galactic distances, as Soft Gamma Repeaters (SGRs) or
anomalous X-ray Pulsars, AXPs. This common Jet nature explain some
connection between GRBs and SGRs spectra. The geometrical
multi-precessing beaming and their reappearance to the observer
may explain the puzzling presence of  X-Ray precursors in GRBs as
well as in SGRs. A mysterious re-brightening afterglows observed
in early and late GRB $030329$ optical transient, like in the $27$
August $1998$ and $18$ April $2001$ SGR $1900+14$ events, might
be  understood as the damped oscillatory imprint left by such a
multi-precessing $\gamma$-X-Optical and Radio  Jet.

\end{abstract}

\section{INTRODUCTION: } Gamma Ray Burst mystery lays  in its
huge energy fluence, sharp variability, extreme cosmic distances
and very different morphology.  A huge isotropic explosion
disagree with the shortest millisecond time
 scales and non-thermal spectra.
  The huge GRBs powers as GRB$990123$ made the final
  collapse of the Fireball  model. New  families of Fire-Ball Jet
  (and their different label names like Hyper-Nova, Supra-Nova, Collapsar )
  models  alleviated the energy budget, by a mild Jet beaming.
   However in this compromise attitude the puzzle of the
   GRB$980425$/SN$1998$bw
    (which require a very thin Jet observed off-axis (Fargion\& Salis 1995),(Fargion1999), has been cured by most  skeptical authors
   by a tenacious cover-up (neglecting or refuting The GRB-SN  existence)
   or claiming the co-existence of an new  zoo of GRBs (Bloom 2000),(Kulkarni).
   These new compromised fountain-like  Fireball model has been collimated in a Jet
 within a $10^o$ angle beam, as a soft link  between past Fireball and emerging
 Jet. However the apparent required GRB power output is still huge ($10^{50}$ erg $s^{-1}$),
  nearly $10^5$ more intense
 than other known maximal explosion power (the Super-Nova one).
  More and more evidences in last years and more recently have shown that
  Super-Nova might  harbor a collimated Jet Gamma Ray Burst
   (GRB$980425$/SN$1998$bw ,GRB$030329$/SN$2003$dh).
   To combine the Super-Nova Luminosity and the apparent huge GRBs
   power one need a very much thinner beam jet, as small as a solid angle $\Delta\Omega/\Omega \simeq$ $10^{-7}$
   or   $10^{-8}$ respect to  $\Omega \simeq 4 \pi$,(corresponding to a Jet  angle  $0.065^o-0.02^o$ ).
    There is a statistical need (Fargion 1999)
   to increase the GRB rate inversely to the beam Jet solid angle.
   The needed SN rate (to explain GRBs) may even exceed the observed one
   (at least SN type Ib andIc) event in our
    observable Universe ($\dot{N_{NS}}$ $\leq 30
   s^{-1}$). Indeed assuming that only a  fraction of the SN
   (with optimistic attitude $0.1$ of all known SN) experience an asymmetric Jet-SN
   explosion, than the  corresponding observed rate  $\dot{N_{GRBs}}$ $\simeq  10^{-5} s^{-1}$ and
   $\dot{N_{SN}}$ $\simeq  3 s^{-1}$  imply
   $\frac{\dot{N_{GRBs}}}{\frac{\Delta\Omega}{\Omega}} \simeq 10^{2} s^{-1} \longleftrightarrow 10^{3} s^{-1}$
   a result  nearly $2-3$  order of magnitude larger  than the observed SN rate.
   In this frame one  must assume a GRB  Jet with a continuous active, decaying  life-time
  much larger than GRBs duration itself at least by a corresponding scale $\tau_{Jet} \simeq 10^{3}\tau_{GRBs}$.
   Indeed we  considered GRBs  (as well as Soft Gamma Repeaters SGR) as
  very thin blazing ($\leq 0.1^o$) precessing Gamma Jets
  spinning and precessing  (Fargion \& Salis 1995,1996,Fargion 1999);
  in this scenario  GRBs are born within a Super-Nova
  power collimated inside a very thin beam  able to blaze  us by an apparent GRB intensities.
\begin{figure}
\plotone{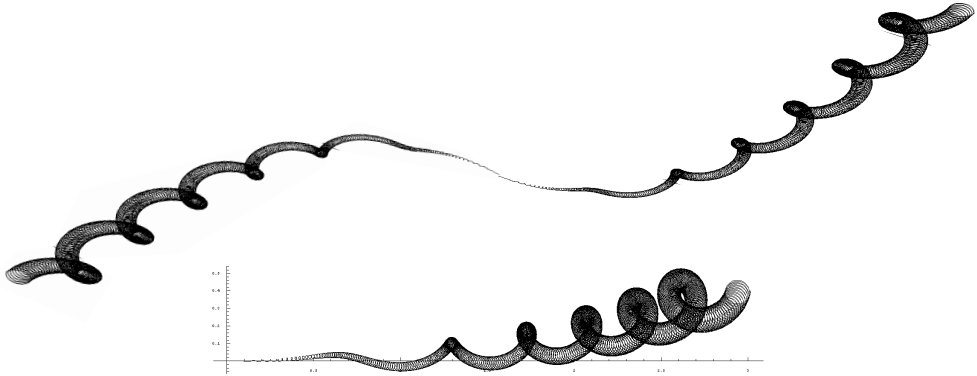}
\caption{\label{fig1}A possible inner
structure 3D of a multi precessing Jet; its cone structures and
its stability at late stage it maybe reflected in the
quasi-periodic repetitions of the  Soft Gamma Repeaters while
beaming to us along the cone edges  toward the observer. Its
early blast  at maximal SN out-put may simulate a brief blazing
GRBs event, while a fast decay (hours scale) may hide its
detectability below the threshold, avoiding in general any
common  GRB  repeater.}
\end{figure}

\begin{figure}
\plotone{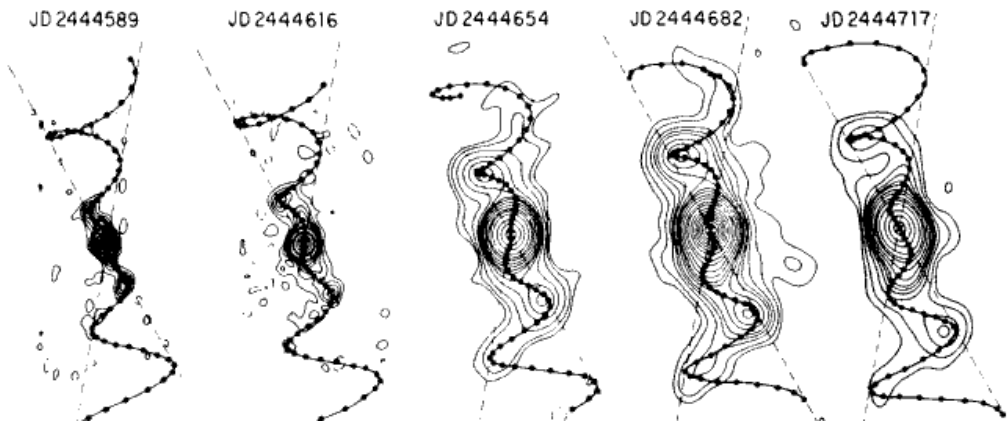} \caption{\label{fig1}The observed structure
3D of a multi precessing Jet in SS433; its structures maybe
reflected in the quasi-periodic repetitions of the  Soft Gamma
Repeaters while beaming to us along the cone edges  toward the
observer.}
\end{figure}
%%%%%%%%%%%%%%%%%%%%%%%%%%%%%%%%%%%%%%%%%%%%%%%%%%%%%%%%%%%%%%%%%%%%%%%%%%%%

The inner angle geometrical dynamics  while spinning and
precessing induce  the wide $\gamma$ burst  variability able to
fit  the very different observed GRBs ones.
 The averaged $\gamma$ jet deflection from the   axis of sight defines a main early power law
decay; an inner damped oscillatory substructure
   may be observed, as the amazing damping oscillatory afterglows in GRB$030329$.  The very thin and
collimated and  long life decaying jet (opening angle $\theta
\leq 0.05^o$, whose decay power law life-time of a few hours
occurs with an exponent $\alpha \simeq -1$), while spinning and
precessing at different scale times, it may trace and may better
explain the wobbling of the $\gamma$ GRBs  and the long train of
damped oscillations of the X tail afterglows   within hours, the
optical transient during days and weeks later. The GRBs
re-brightening are no longer a mystery as in a one-shoot model.
These wobbling signatures may be also be found in rarest and most
powerful and studied SGRs events. The spread and wide conical
shape of these precessing twin jets may be recognized in a few
relic SNRs as in  the twin SN 1987A wide external  rings, the
Vela arcs and the spectacular Egg Nebula dynamical shape.
\subsection{The  geometrical multi-precessing Gamma Jet by I.C. in GRB }
 We imagine the GRB and SGR
nature as the early and the late stages of jets fueled first by
SN event and later by an asymmetric accretion disk or a companion
(WD, NS) star.
%%%%%%%%%%%%%%%%%%%%%%%%%%%%%%%%%%%%%%%%%%%%%%%%%%%%%%%%%%%%%%%%%%%%%%%

%%%%%%%%%%%%%%%%%%%%%%%%%%%%%%%%%%%%%%%%%%%%%%%%%%%%%%%%%%%%%%%%%%%%%%%%

The binary angular velocity $\omega_b$ bends the angle as
$\theta_1(t) = \sqrt{\theta_{1 m}^2 + (\omega_b t)^2}$ or more
generally a multi-precessing angle
$\theta_1(t)$:$\theta_1(t)=\sqrt{\theta_{x}^2+\theta_{y}^2 } $
\\
 $ \theta_{x}(t) =                               %\theta_{1 m}+
  \theta_{b} sin(\omega_{b} t+ \varphi_{b} )+
  \theta_{psr}sin(\omega_{psr} t +  \varphi_{psr})+
  \theta_{N}sin(\omega_{N} t  + \varphi_{N})$
\\
$\theta_{y}(t) = \theta_{1 m}+
  \theta_{b} cos(\omega_{b} t + \varphi_{b})+
  \theta_{psr} cos(\omega_{psr} t +  \varphi_{psr})+
  \theta_{N} cos(\omega_{N} t  + \varphi_{N})$
  \\
where $\theta_{1 m}$ is the minimal impact angle parameter of the
jet toward the observer, $\theta_{b}$, $\theta_{psr}$,
$\theta_{N}$ are, in the order, the maximal opening precessing
angles due to the binary, spinning pulsar, nutation mode of the
multi-precessing jet axis. The arbitrary phase $ \varphi_{b}$, $
\varphi_{psr}$, $\varphi_{N}$, for the binary, spinning pulsar
and nutation,  are able to fit the complicated GRBs flux
evolution in most GRB  event scenario. Naturally it is very
possible to enlarge the parameter to a fourth precession angular
component whose presence may better fit the wide spread of scale
variability; here we shall constrains to a three parameter
precession  beam.
%%%%%%%%%%%%%%%%%%%%%%%%%%%%%%%%%%%%%%%%%%%%%%%%%%%%%%%%%%%

%%%%%%%%%%%%%%%%%%%%%%%%%%%%%%%%%%%%%%%%%%%%%%%%%%%%%%%%%%%%%%%%%

%%%%%%%%%%%%%%%%%%%%%%%%%%%%%%%%%%%%%%%%%%%%%%%%%%%%%%%%%%%%%%%%%%%%%%%%%%%%%%%%%%%%
\begin{figure}\plottwo{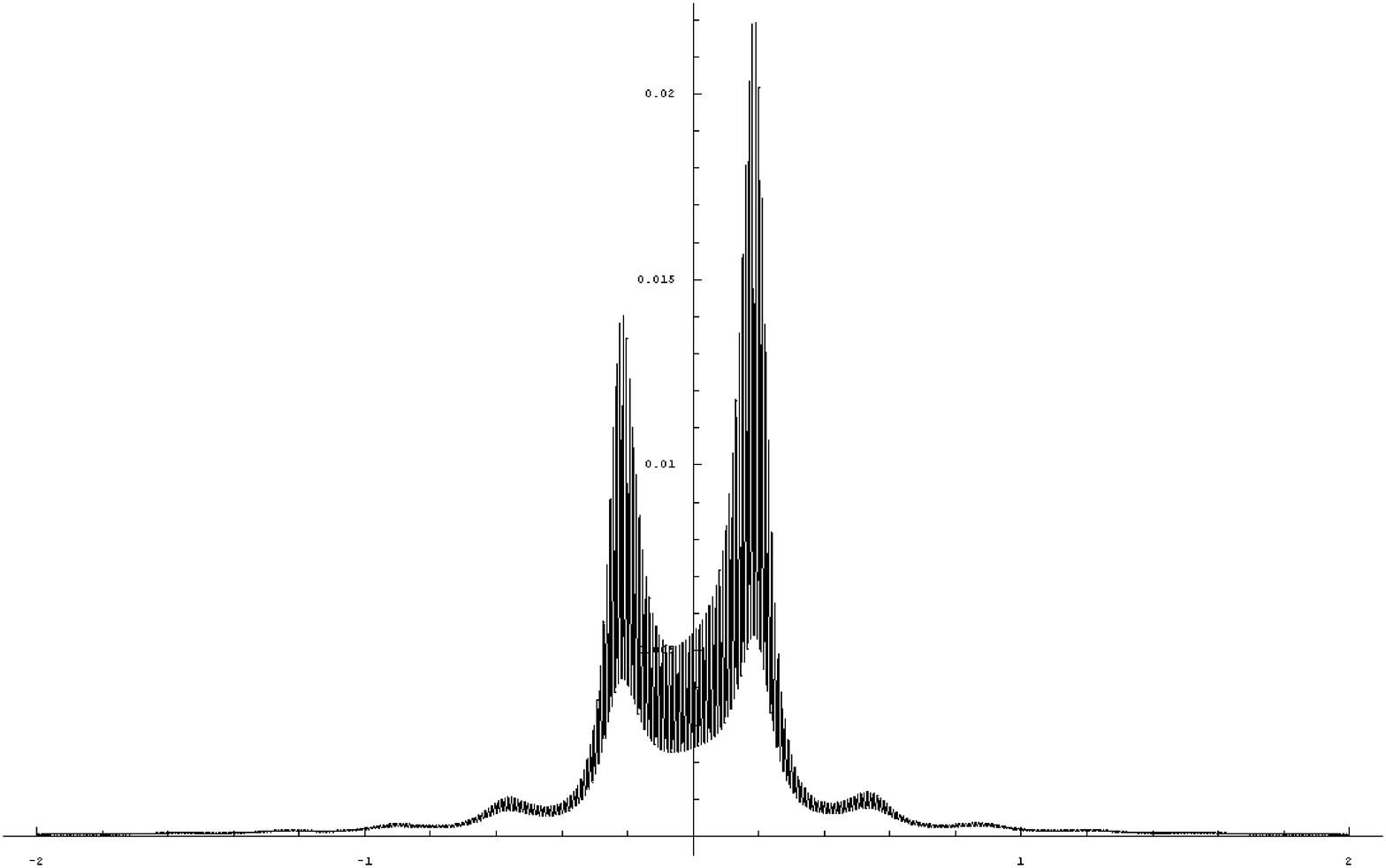}{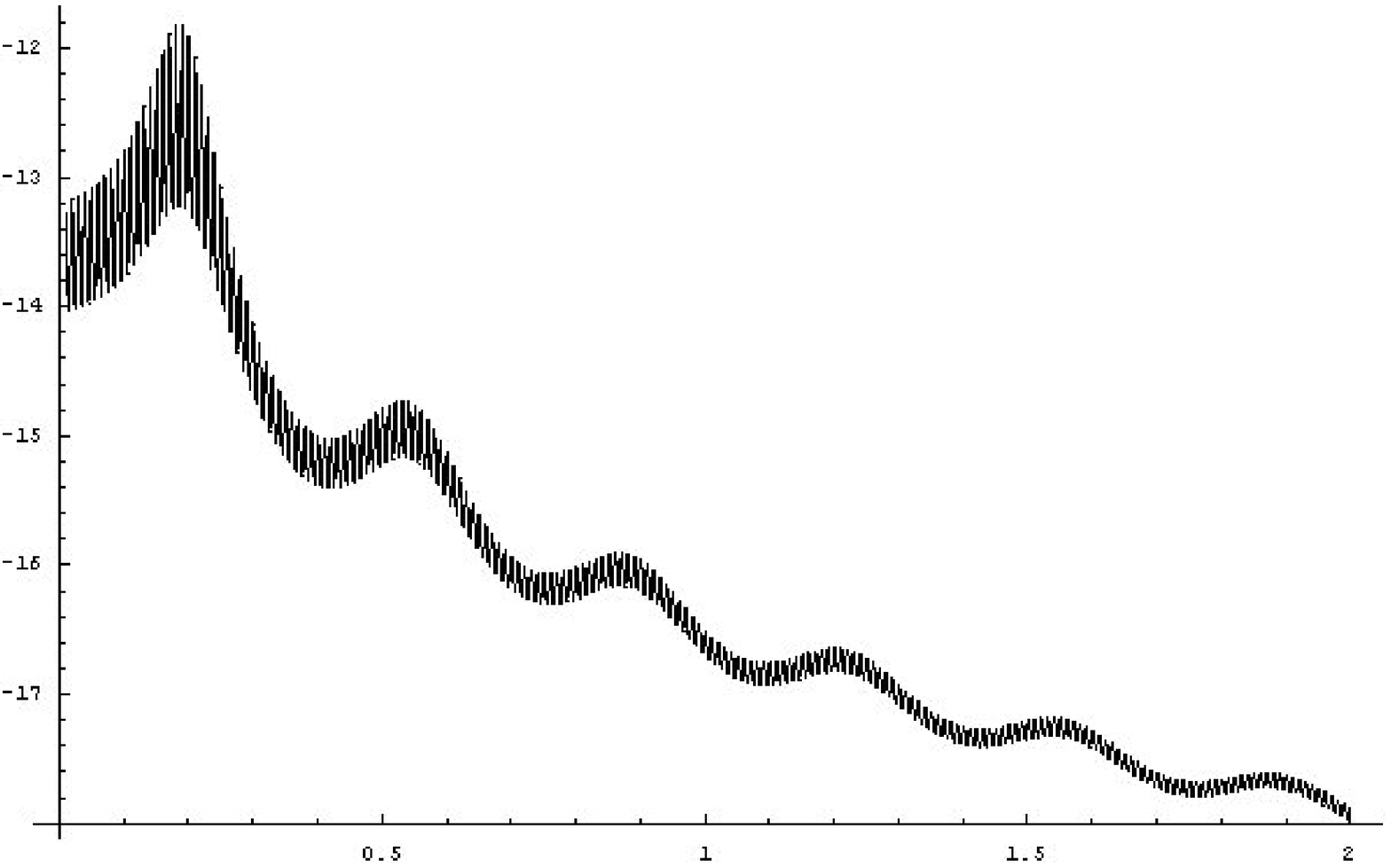} %for centering: act on hspace argument
\caption{ Left: Multi bump afterglow behaviour of the intense
precessing Jet above whose blazing shows the characteristic
oscillatory damped decay as the recent GRB $030329$ and the
intense SGR on $27$ August 1998. The luminosity starting time is
assumed near zero (at SN birth time). In present simulation the
assumed Lorents factor is $\gamma_e$$= 2 \cdot 10^3$. Right: Semi
Log. Multi bump Flux Intensity behaviour in linear time scales,
normalized to visual magnitude for the previous precessing Jet
simulating the characteristic oscillatory damped decay as the
recent GRB $030329$ and the intense SGR$1900+14$ on $27$ August
1998; time scale are arbitrary; in the GRB $030329$ the unity
corresponds to nearly day scale while in SGR event the unity in
much smaller tens seconds scale }
\end{figure}
%%%%%%%%%%%%%%%%%%%%%%%%%%%%%%%%%%%%%%%%%%%%%%%%%%%%%%%%%%%%%%%%%%%%%%%%%%%%%%%%%%%

%%%%%%%%%%%%%%%%%%%%%%%%%%%%%%%%%%%%%%%%%%%%%%%%%%%%%%%%%%%%
%%%%%%\begin{figure}\centering\includegraphics[width=9cm]{LogFlux.eps} %for centering: act on hspace argument
%%%%%%\caption { Multi bump Flux Intensity behaviour, normalized to
%%%%%%visual magnitude for the previous  precessing Jet showing the
%%%%%%characteristic oscillatory damped decay as the recent GRB
%%%%%%$030329$ and the intense SGR on $27$ August 1998 }
%%%%%%\end{figure}

    For a 3D pattern and its projection along the vertical
axis in an orthogonal 2D plane see following descriptive pictures.
The different  angular velocities are combined in the
multi-precession wobbling. Each bending component  is keeping
memory either of the pulsar jet spin angular velocity
($\omega_{psr}$) and its opening angle $\theta_{psr}$, its
nutation speed ($\omega_N$) and nutation angle $\theta_{N}$ (due
to possible inertial momentum anisotropies or beam-accretion disk
torques); a slower  precession  by the binary $\omega_b$
companion (and its corresponding open angle $\theta_{b}$) will
modulate the overall jet precession. On average, from eq.(3) the
$\gamma$ flux and the $X$, optical afterglow are decaying on time
as $t^{-\alpha}$ , where $\alpha \simeq 1-2$; however the more
complicated spinning and precessing jet blazing is responsible
for inner small scales wide morphology of GRBs and SGRs as well
as their partial internal periodicity. The consequent $\gamma$
time evolution and spectra derived in this ideal models may be
compared successfully with observed GRB data evolution.

%%%%%%%%%%%%%%%%%%%%%%%%%%%%%%%%%%%%%%%%%%%%%%%%%%%%%%%%%%%%%%%
\begin{figure}\plottwo{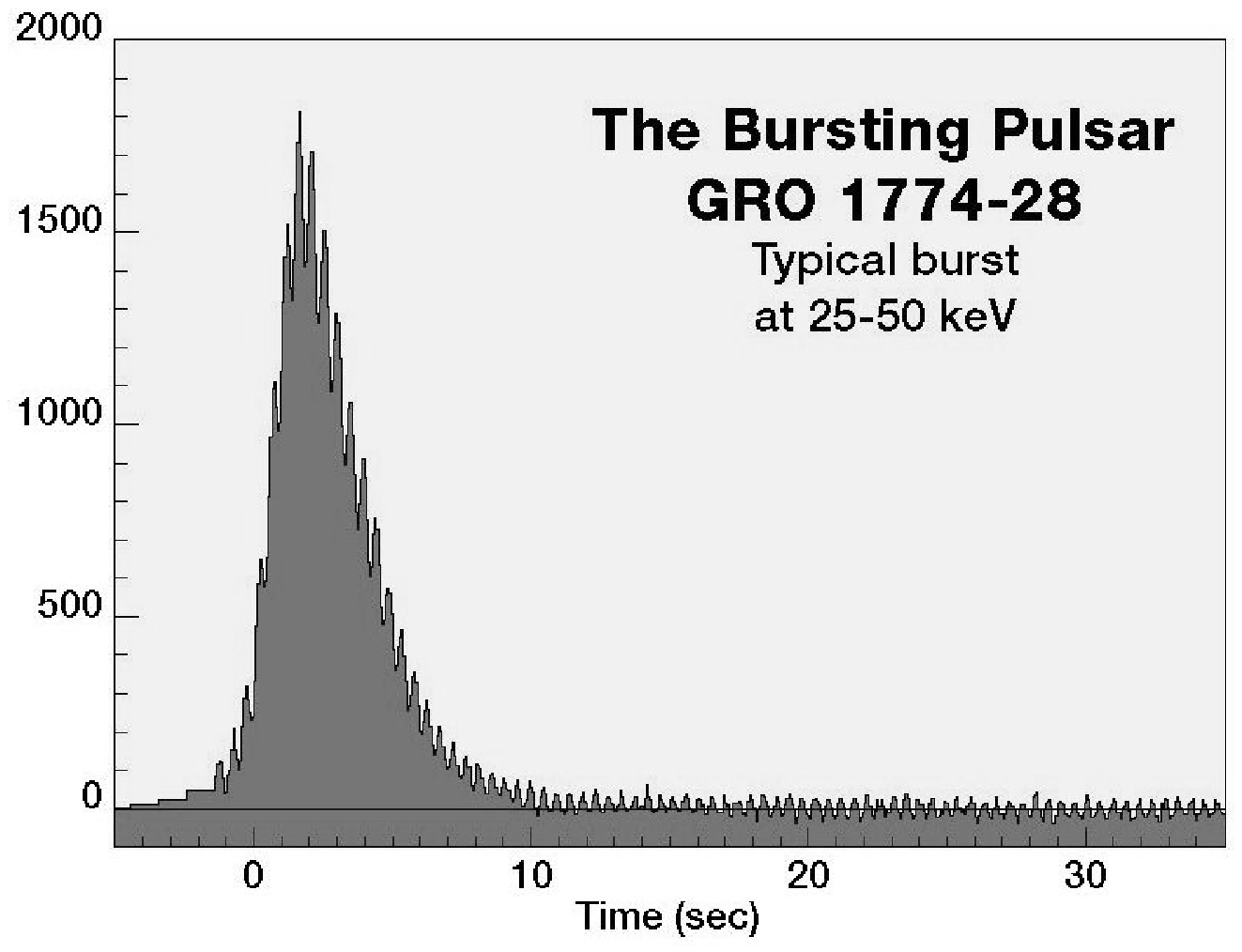}{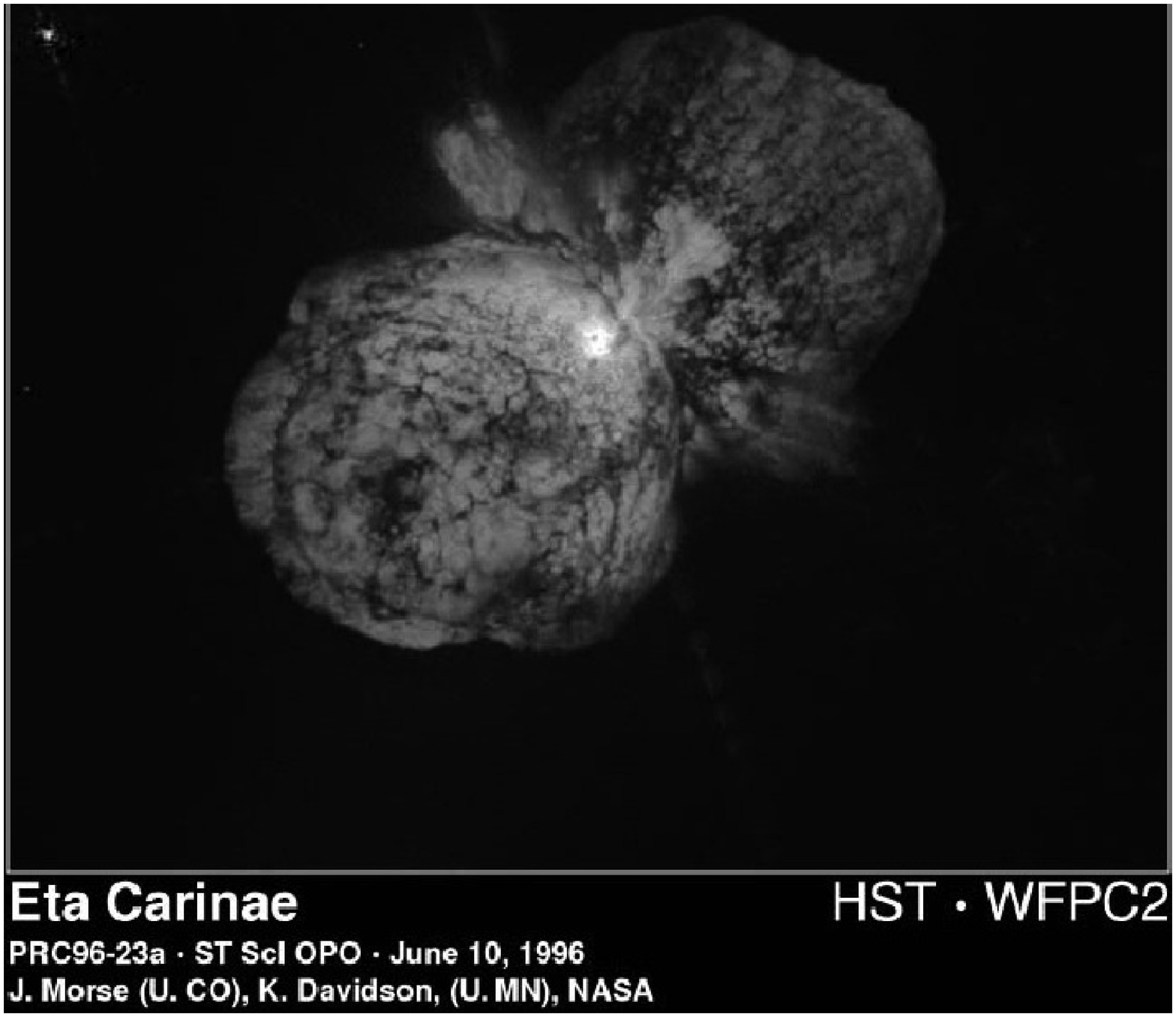}
\caption {Gamma Luminosity of GRO17174-28; its peculiar
variability might be related to a semi-periodic flashing of a
precessing Jet. Eta Carina Nebula that very probably hosts a
precessing Jet. Such a mysterious hour-glass may hide in the
center of the nebula whose presence is  evident by the help of a
diffused nearby cloud target as well as by thin jet fingers and
mysterious luminosity variability observed in the past century }
\end{figure}
%%%%%%%%%%%%%%%%%%%%%%%%%%%%%%%%%%%%%%%%%%%%%%%%%%%%%%
The $\gamma$ Jet is born mainly by Inverse Compton Scattering by
GeVs electron pairs onto thermal photons (Fargion 1995-96-98-99)
in nearly vacuum space. Therefore these   electron pairs are
boosted in the Jet at Lorentz factor $\gamma_e \geq 2 \cdot 10^3$.
   Their consequent Inverse Compton Scattering will induce a parallel $\gamma$ Jet
   whose beam angle is $\Delta \theta \leq \frac{1}{\gamma} \simeq 5\cdot10^{-4} rad \simeq 0.0285^o $
   and a wider, less collimated X, Optical cone. These beaming
   angles agree with the one assumed to explain the required beamed GRBs-SN
   powers.   Indeed  the electron pair Jet may generate  a secondary beamed synchrotron
   radiation component at radio energies, in analogy to the behaviour
    of  BL Lac blazars whose hardest TeV $\gamma$ component is made by Inverse Compton Scattering
    while its  correlated  X band emission is due to the synchrotron component. Anyway the inner jet
   is dominated by harder photons while the external cone contains softer $X$,
   optical and radio waves. A jet angle related by a relativistic kinematics would imply $\theta \sim
\frac{1}{\gamma_e}$, where $\gamma_e$ is found to reach $\gamma_e
\simeq 10^3 \div 10^4$ (Fargion 1999). At first approximation the
gamma constrains is given by Inverse Compton relation: $<
\epsilon_\gamma > \simeq \gamma_e^2 \, k T$ for $kT \simeq
10^{-3}-10^{-1}\, eV$ and $E_e \sim GeVs$ leading to
characteristic X-$\gamma$ GRB spectra.  The origin of $GeVs$
electron pairs are  very probably decayed secondary related to
primary inner muon pairs jets, able to cross dense stellar target
(Fargion 2003).
%%%%%%%%%%%%%%%%%%%%%%%%%%%%%%%%%%%%%%%%%%%%%%%%%%%%%%%%%%%%%%%%%%%%
\section{Conclusions: Neutrino-Muon Jets Progenitors for Gamma-X-Burst}
The same GeVs electron pair Jet may generate  a secondary beamed
synchrotron   radiation component at radio energies, in analogy
to the behaviour    of  BL Lac blazars whose hardest TeV $\gamma$
component is made by Inverse Compton Scattering
    while its  correlated radio bumps  is due to the synchrotron component. Anyway the inner jet
   is dominated by harder photons while the external cone contains softer $X$,
   optical and radio waves.  The very peculiar oscillating GRB$970508$
    optical variability did show  both a re-brightening and
    remarkable multi-bump variability in radio wave-lenght.  For this reason and
    we  believe that this fluctuations were indeed to be related to the Jet
    precession and not to any interstellar scintillation. There is not any direct
   correlation between the $\gamma$ Jet made up by I.C. scattering and the Radio Jet because
   the latter is dominated by the external magnetic field energy density: there
   maybe a different beaming opening and a consequent different time modulation
   respect to the inner $\gamma$ Jet. However the present wide
   energy power emission between SN2002ap and GRB$030329$ radio light curves
   makes probable a beaming angle comparable: $\leq 10^{-3}- 10^{-4}$ radiant.

  These GRBs Jets are originated by NSs
or BH in binary system or disk powered by infall matter; their
relics (or they progenitors) are nearly steady X-ray Pulsars
whose fast blazing is source of SGRs. This external $\gamma$ Jet
has a chain of progenitor identities: it is very probably
originated by a very collimated inner primary muon Jet pairs at
TeVs-Pevs energies. These muons could cross with negligible
absorption the dense target lights along the SN explosions,
nearly transparent to photon-photon opacities. The high energy
relativistic muons (tens TeVs-PeVs energies) decay in flight in
electron pairs where the baryon density is still negligible;
these muons are source, by decay in flight to Tevs-GeV electron
pair showering whose final Inverse Compton Scattering with nearby
thermal photon is the final primary of the observed hard $X$ -
$\gamma$ Jet. The cost of this long chain of reactions  is a poor
energy conversion, but the benefit is the possibility to explain
the $\gamma$ escape from a very dense explosive and polluted (by
matter and radiation) narrow  volume.
 Its inner Jet ruled by relativistic Inverse Compton Scattering, has the hardest and rarest beamed
GeVs-MeVs photons (as the rare and long $5000$ s  life  EGRET
GRB$940217$ one) but its external Jet cones are dressed by softer
and softer photons. The complex variability of GRBs and SGRs are
simulated successfully by the equations above (Fargion \& Salis
1995,1996,1998, Fargion 1999); the consequent geometrical beamed
Jet blazing may lead also to the observed widest morphology
$X-\gamma$ signatures and rapid re-brightening as X-ray
precursors. The mystery therefore is   not longer in an apparent
huge GRB luminosity, but in an extreme beam jet
   collimation and precession.

\end{document}